# Human-Centered AI Maturity Model (HCAI-MM): An Organizational Design Perspective

Stuart Winby, Palo Alto, California, USA, Stu@winby.biz

Wei Xu, HCAI Labs, California, USA, weixu6@yahoo.com, https://orcid.org/0000-0001-8913-2672

*Abstract* - Human-centered artificial intelligence (HCAI) is an approach to AI design, development, and deployment that prioritizes human needs, values, and experiences, ensuring that technology enhances human capabilities, well-being, and workforce empowerment. While HCAI has gained prominence in academic discourse and organizational practice, its implementation remains constrained by the absence of methodological guidance and structured frameworks. In particular, HCAI and organizational design practices are often treated separately, despite their interdependence in shaping effective socio-technical systems. This chapter addresses this gap by introducing the Human-Centered AI Maturity Model (HCAI-MM)**,** a structured framework that enables organizations to evaluate, monitor, and advance their capacity to design and implement HCAI solutions. The model specifies stages of maturity, metrics, tools, governance mechanisms, and best practices, supported by case studies, while also incorporating an organizational design methodology that operationalizes maturity progression. Encompassing dimensions such as human-AI collaboration, explainability, fairness, and user experience, the HCAI-MM provides a roadmap for organizations to move from novice to advanced levels of maturity, aligning AI technologies with human values and organizational design principles

## 1. Introduction

Artificial Intelligence (AI) refers to the simulation of human intelligence processes by machines, particularly computer systems, enabling them to learn, reason, and make decisions. There are two main ways AI are being used in organizations: substituting humans through automation and augmenting humans' ability to search, comprehend problems, generate and evaluate solutions, and make choices. This chapter focuses on augmentation, the intentional collaboration of humans with smart technologies, referred to as HCAI (Xu, 2019; Shneiderman, 2020).

HCAI refers to the design, development, deployment, and use of AI systems that prioritize human needs, values, and experiences, enhancing human capabilities rather than replacing them. This is contrary to AI strategies, which decompose processes into tasks, automate as many tasks as possible, and what is left over is combined into new roles and responsibilities. HCAI emphasizes ensuring that technology serves the needs and values of people, rather than merely automating tasks. This approach emphasizes collaboration between humans and AI, ensuring that technology enhances human capabilities and decision-making while promoting transparency, fairness, and ethical considerations (Schmager, Pappas, & Vassilakopoulou, 2023). By fostering collaboration between humans and AI, this approach impacts the design of an organization by empowering individuals, decentralizing information, decision-making, creativity, and productivity through self-managing structures and processes (Snow & Fjeldstad, 2024).

HCAI practices encompass a set of methodologies and strategies that prioritize the involvement, needs, and well-being of users in the design, development, deployment, and use of AI systems. These practices involve engaging stakeholders throughout the AI lifecycle, from ideation and design to testing and implementation, ensuring that the technology is intuitive, accessible, and aligned with ethical considerations (Shneiderman, 2020).

Organizational capabilities are the foundational skills and resources that enable the implementation of HCAI practices, while HCAI practices are the actual methodologies or actions taken based on those capabilities. The relationship between HCAI practices and organizational design is critical, as organizations must align their structures, cultures, and processes to support and implement these practices effectively (Daugherty & Wilson, 2018). For instance, the organizational capabilities, such as a strong machine learning engineering talent pool and organization infrastructure, serve as the bedrock for HCAI applications like AI-powered code completion and bug detection. Without a team of skilled engineers, adequate computational resources, high-quality data, and a culture that fosters innovation and collaboration, the organization would struggle to develop and deploy AI tools that significantly enhance software development efficiency. This alignment not only drives technological advancements but also ensures that the solutions effectively cater to user needs and improve overall developer productivity.

HCAI practices are also essential to the competitiveness and performance of today's state-of-the-art modern organizations, as they foster a human-centered approach that drives innovation and responsiveness in technology development. Given technology is changing at a faster rate than organizations, HCAI practices will allow organizations to change and reconfigure at a rate commensurate to technology change (Pasmore, Winby, Mohrman, & Vanasse, 2018). By integrating HCAI practices into operational frameworks, organizations can enhance their agility through innovation, differentiate themselves in the market, and improve overall performance by delivering AI solutions that effectively meet the evolving needs of users and stakeholders.

This chapter aims to explore the integration of HCAI practices within today's modern organizational design by introducing an HCAI-MM that assesses an organization's progress in adopting these practices. It examines how HCAI is incorporated into the design of organizations. The chapter provides a structured approach to evaluate and enhance AI initiatives, with an emphasis on driving competitiveness and performance while fostering ethical and impactful AI outcomes. This integration serves as a strategic framework for organizations seeking to elevate their HCAI maturity, ensuring that technology not only aligns with human values and societal well-being but also enhances their market position and responsiveness to stakeholder needs.

## 2. Human-Centered AI Practice

HCAI practice is an approach that prioritizes the needs, values, and experiences of people in the design, deployment, and use of AI systems, ensuring that technology enhances human well-being and promotes ethical outcomes. HCAI practices emphasize collaboration between AI developers and end-users to ensure that technology enhances human and organizational capabilities and addresses societal challenges effectively. Table 1 provides an overview of key HCAI practices and their potential organizational impact.

**TABLE 1 HCAI Practices and Organizational Impact**

| Key HCAI Practices | Description | Impact |
|---|---|---|
| **User Research and Engagement** | Involve end-users early in the development process through interviews, surveys, usability testing, and deliberation design shops. Equip users with the knowledge and tools to understand and interact with AI systems | Ensures that AI systems are relevant and meet user needs, leading to higher adoption rates and satisfaction. User interaction with AI in deliberation design impacts new work and organization designs. |
| **Iterative Design** | Utilize rapid prototyping and iterative feedback loops to refine AI solutions continuously. | Enhances product quality and reduces the risk of failures by incorporating user feedback early and often. |
| **Ethical Alignment** | Adopt ethical guidelines that prioritize fairness, transparency, and accountability in AI systems. This involves conducting ethical assessments and considering societal implications of deploying AI systems. | Builds trust with users and stakeholders, minimizing reputational risks and aligning with regulatory requirements |
| **Explainability and Transparency** | Develop AI systems that can explain their decisions in understandable terms. Explain what decisions are made, what data is used, and how outcomes are achieved. | Increases user trust and reduces resistance to AI adoption, especially in sectors like healthcare and finance, where decisions have significant consequences. |
| **Safety and Reliability** | Ensure AI systems are safe and reliable, minimizing the risk of unintended consequences or failures. This includes testing, validation, and ongoing monitoring. | AI systems that are safe and reliable |
| **Cross-Disciplinary Collaboration**: | Involve stakeholders /experts from various fields (e.g., ethicists, sociologists, designers) alongside technologists to ensure a well-rounded perspective and solutions | Results in more robust solutions that take into account diverse perspectives, potentially leading to innovative applications and reduced biases. |
| **Focus on Empowerment**: | Design AI systems that augment human capability rather than replace it. AI systems that empower users and enhance their abilities rather than replacing them. This typically entails empowering through information, skills, and decision-making. | Promotes user empowerment and job satisfaction, while also enhancing productivity and creativity within team-based work systems. Typically changes work design and org structure. |
| **Continuous Learning and Adaptation** | Implement mechanisms for AI systems to learn from user interactions and adapt over time. Implement mechanisms for ongoing evaluation and feedback, allowing for iterative improvements. | Creates dynamic work systems that remain relevant and effective, leading to sustained user engagement and improved outcomes |

2.1 HCAI Practice Challenges

Current HCAI practices face challenges due to the absence of a comprehensive maturity model, which can hinder organizations from effectively assessing and improving their AI systems (Deloitte,2020; Hartikainen,Vaananen, Olsson, 2023; Wilkens, Langholf, Ontruo & Kluge, 2012). Additionally, organizations may struggle to identify best practices and scalable strategies for enhancing user experience, resulting in inconsistent implementation of human-centered approaches. Implementing a maturity model will enable organizations to systematically evaluate their practices, identify weaknesses, and adopt best practices, ultimately fostering enhanced transparency, user engagement, and ethical considerations in AI deployment (Sonntag, Mehmann, Mehmann, & Teuteberg, 2024).

## 3. HCAI Maturity Model (HCAI-MM): Overview

Maturity implies progress in the development of a system to reach its target state. Typically, maturity is considered a gradual process, consisting of multiple stages that are staggered sequentially, each specifying the requirements for that level of complexity. A maturity model defines a sequence of maturity levels and for each level describes an expected, desired development path in successive discrete stages. Thus, a maturity model can show how an organization's capabilities progressively evolve along an expected logical maturity path. The degree of maturity can be measured qualitatively and quantitatively in a continuous manner (Hein-Pensel, Winkler, Bruchner & Others, 2023).

The HCAI-MM is a framework designed to help organizations assess and enhance their capabilities in developing and deploying AI systems that prioritize human values and needs. It emphasizes a structured approach to integrating ethical considerations, user experience, and stakeholder engagement into AI initiatives, allowing organizations to evaluate their current AI practices, identify areas for improvement, and guide the responsible evolution of their AI strategies. By progressing through various maturity levels, organizations can ensure that their AI solutions are not only technologically advanced but also align with human values and promote positive outcomes for users and stakeholders.

Employing HCAI-MM has multiple organizational benefits. HCAI-MM equips organizations with tools, metrics, and methodologies for assessing their current maturity level regarding HCAI practices. The HCAI-MM helps organizations to systematically enhance their AI maturity, ensure alignment with business objectives, and foster a culture of continuous learning and innovation in AI initiatives.

## 3.1 HCAI Guiding Principles

Human-centered design principles serve as the guiding philosophy for creating user-centered solutions, while the HCAI-MM provides a framework for evaluating how well organizations integrate these principles in their AI initiatives. See Table 1 below of HCAI Guiding Principles (Xu, Gao, & Dainoff, 2024).

**TABLE 2: HCAI Guiding Principles**

| # | Guiding Principle | Definition and HCAI Goals |
|---|---|---|
| 1 | Transparency and Explainability | Provide explainable and understandable AI output for users to enhance human trust and empower informed decisions. |
| 2 | Human control and empowerment | Allow humans to understand, influence, and control AI behavior when necessary; Ensure AI aligns with human needs. |
| 3 | Ethical Alignment | Develop AI in alignment with ethical and societal norms to preserve human values and privacy, foster trust, and minimize harm. |
| 4 | User Experience | Create interactions that are engaging, intuitive, accessible, and aligned with user expectations. |
| 5 | Human-Led Collaboration with AI | Design AI to enhance human abilities and human-led collaboration to enhance productivity and effectiveness. |
| 6 | Safety and Robustness | Prioritize human safety and maintain reliability in diverse scenarios to ensure resilience and reduce potential risks to humans. |
| 7 | Accountability | Ensure responsible AI mechanisms for accountability to hold humans (e.g., operators) accountable for AI actions. |
| 8 | Sustainability | Develop AI to support environmental, social, and economic well-being to prioritize human well-being and cultivate resilient ecosystems while aligning with sustainability. |

## 3.2 HCAI-MM and Organizational Design

During the last decade, AI has made dramatic inroads into the operations of organizations and is becoming a mainstream organizational technology (Daughterty & Wilson, 2018). As mentioned earlier, two main ways AI is being used in organizations: substituting for humans on routine tasks and augmenting human's ability in complex situations to search, comprehend problems, generate and evaluate alternative solutions, and make choices. Organizations are increasingly incorporating AI into their operations, both through automation and human-centered AI, to make decisions and act to be better able to solve problems and adapt faster and more effectively to changes in their environment. AI's influence on how activities and resources are organized continues to increase (Shrestha, Ben-Menahem, & von Krogh, 2019).

HCAI and organizational design are intertwined because AI reshapes how work is done and how organizations function. At the same time, changes to organizational design are necessary to integrate and optimize HCAI systems.

An understanding of the HCAI-MM can help organizations identify necessary changes in structure, culture, processes, and strategies to leverage AI technologies, enhancing the overall effectiveness of the organization. Likewise, the effectiveness of HCAI is to a large extent a function of its organization design, again defined through HCAI-MM. An organization's design must support the human-centered principles needed for high HCAI maturity, and the HCAI maturity assessment can provide insights to improve organizational design (Wilkens, Lupp, & Langholf, 2023).

The relationship between an HCAI-MM and organizational design can be understood through several organizational processes, as shown in Table 3.

**TABLE 3 HCAI Maturity Model and Organization Design**

| Organization Processes | Description |
|---|---|
| **Organizational Readiness** | Understanding where an organization is positioned on the HCAI-MM helps leaders gauge their readiness to implement HCAI. This involves assessing existing structures, processes, culture, skills, and technologies that shape organizational design. (See Table 4 – Readiness Assessment) |
| **Resource Allocation** | Organizations at different maturity stages will have distinct requirements regarding resource allocation—both in terms of budgeting and human capital. A mature HCAI organization might have dedicated teams and roles for HCAI, while a start-up organization may need to upskill existing employees or hire new talent. |
| **Governance and Strategy** | An HCAI-MM emphasizes the importance of strategic alignment. As organizations evolve in their HCAI capabilities, their governance structures may need to adapt accordingly. Effective organizational design ensures that decision-making processes and workflows are structured to support HCAI initiatives and the necessary oversight for ethical and effective use of AI technology. |
| **Integration of AI in Business Strategy** | Organizations that effectively embed HCAI into their strategic framework tend to demonstrate enhanced effectiveness. This involves redesigning organizational structures to ensure that HCAI initiatives are aligned with business objectives and that there is collaboration across departments for data-sharing and decision-making. |
| **Impact on Performance Metrics** | The stages of HCAI maturity are likely to reflect on how organizational effectiveness is measured. As organizations adopt more advanced AI solutions, performance metrics may shift from traditional KPIs to those that assess HCAI-specific impacts, such as improved decision-making speed or enhanced employee experience. |
| **Process Optimization** | At higher levels of maturity, organizations tend to optimize their processes through HCAI insights. As a result, organizational work design may need to change to account for more automated and data-driven processes, which can lead to greater efficiency and effectiveness. |
| **Feedback Loops** | As organizations advance in HCAI maturity, their processes and organizational design will benefit from the integration of continuous feedback loops. This means that information gathered from AI applications will inform strategic and operational adjustments, facilitating a more agile organization. |
| **Cultural Transformation** | Advancing through an HCAI maturity model often requires a cultural shift within the organization. This includes fostering a culture of innovation, agility, and risk-taking. Organizational effectiveness is closely tied to how well the culture supports HCAI initiatives. |

### 3.2.1 Augmenting Organization Problem-Solving

Compared to tasks that can be standardized, tasks that require complex problem solving, creativity, or emotional intelligence are less amenable to automation. For those types of tasks, AI is used to augment human problem-solving and decision-making. HCAI problem-solving and decision-making typically improve the comprehension of humans to understand their external and internal environments. Comprehension is challenging when operating in complex, dynamic environments characterized by a large number of constantly changing, interacting elements. The increased availability of data captured from a variety of sources makes it possible to use algorithms to improve organizational comprehension (Schmidt, 2017). Augmenting organization problem solving is a key organizational capability central to innovation, performance improvement, and adaptability.

### 3.2.2 Key HCAI Practice: Deliberations

A deliberation is a structured discussion or process where a group of people carefully considers and weighs the pros and cons of different options, solutions, or courses of action related to a specific issue. The goal of a deliberation is to reach a clear understanding of the problem and a well-reasoned and informed decision, typically involving a variety of perspectives and evidence (Pava, 1983).

As a HCAI practice and a socio-technical design method, deliberations facilitate user involvement in the design and development of HCAI systems. This includes consulting with users, soliciting feedback on design choices, and incorporating human perspectives into the decision-making process. The deliberative process aims to make sure that human control over the AI system is maintained.

In organization design terms, the deliberation is an organizational unit of analysis of human-machine collaboration and an iterative process of decision-making, typically applied to the design, development, and deployment of HCAI systems. Outputs of deliberations frequently result in changes in the organization's work design. Deliberations constitute an effective unit of analysis: inputs, conversion, and outputs across individual, group, cross-functional, ecosystem, and customer boundaries. Deliberations are a form of work and organization design typically configured in a socio-technical design lab management process (Winby & Worley, 2014).

### 3.2.3 Improving Organizational Choice

Deliberation, as an HCAI practice, improves organizational choice. In confronting organizational problems and/or opportunities, generating options involves problem-solving activities that seek to develop customized courses of action. Choice is the process of predicting the consequences of each of the options, evaluating options across the option sets, and selecting and implementing a course of action. Organizational choice refers to the decisions made by organizational actors, particularly managers and leaders, that shape the structure, processes, and culture of the organization. It's about intentionally selecting specific configurations and arrangements of resources and activities to achieve organizational goals. HCAI, and specifically the deliberation process, allows for continuous organizational choice and reconfigurability as a dynamic organizational capability.

### 3.2.4 Implications for Organizational Design

As AI continues to make advances, managers will increasingly incorporate AI into their organizations, either through automation or HCAI. It is important to keep "humans-in-the-loop" to ensure that AI performs according to the organization's requirements and can be adapted as needed (Hermann, & Pfeffer, 2023; Gronsund & Aanestad,2020). The key to keeping humans actively in the loop is to design the organization and technology such that humans maintain supervisory control over AI. This requires that human actors understand the actions of their AI machine collaborators as well as the reasons why the actions are taken.

To be effective moving forward, organizations will need to position AI as a strategic resource. The organization design challenge for HCAI is to create organizational environments in which humans and AI can collaborate effectively. Managers must learn and diffuse HCAI practices that are conducive to collaboration in their organizations. Ultimately, the success of their efforts rests on having humans who have the knowledge and skill to work effectively with their AI colleagues as well as the processes and infrastructures, like the socio-technical systems design lab management process, that allow humans and AI to communicate and collaborate (Fjeldstad, Snow, Miles, Lettl, 2012).

### 3.2.5 Socio-technical Systems (STS)

Modern organization design has two streams of design approaches. One stream, strategic organization design, is a top-down design process that begins with strategy and is applied at the enterprise, business unit, geographical, and functional levels. The second stream, the socio-technical systems (STS) approach focuses on the alignment of technology involved in doing the work and the social organization design created to perform that work (Galbraith, 2014).

STS has a long tradition of theory development and practical application in understanding the integration of humans and technology (Trist & Murray, 1990). The core premise of STS is that organizations function as joint systems composed of social and technical elements, which must be designed and optimized together. Social elements include people, culture, and work practices, while technical elements encompass tools, processes, and technologies. If one system is prioritized at the expense of the other, the result is often inefficiency, reduced effectiveness, or worker dissatisfaction (Trist & Murray, 1990).

A central contribution of STS theory to organization design is the division of labor and the concept of the *work system*. Unlike viewing an organization merely as a collection of jobs, people, or technologies, a work system is treated as a design unit - a division of labor that functions as an integrated system. A work system: (1) is organized to accomplish a defined purpose that provides direction for its activities; (2) consists of interdependent elements such as people, tasks, technologies, structures, and environments; and (3) operates as a system with inputs, a transformation process, outputs, boundaries, and feedback mechanisms for monitoring and self-regulation. Variances - deviations from goals or specifications - are addressed through adjustments in organization design and/or technology, enabling high performance, reconfigurability, and adaptability. These properties allow the work system to remain stable, adaptive, and aligned with its purpose (Trist & Murray, 1990).

STS has also introduced foundational concepts such as quality of working life, participative design, and human well-being as integral to STS theory and practice. Organizations consistently face the dual challenge of achieving efficiency, innovation, and adaptability (productivity and agility) while ensuring that work remains humane, meaningful, health-promoting, and conducive to a high quality of work life. At the core of this balance lies the division of labor—how tasks are distributed, who has autonomy, and who exercises control over different parts of the system. The introduction of digital and AI tools frequently reshapes these dynamics, sometimes in subtle ways, making it essential that such shifts are anticipated and deliberately designed. These principles are also central to the philosophy and goals of human-centered AI (Govers & Van Amelsvoort, 2023).

### 3.2.6 STS and HCAI

HCAI and STS are related in that HCAI is a socio-technical system. HCAI and STS are both oriented to achieve joint optimization between humans and technology. Both HCAI and STS are concerned with designing organization systems that are effective, efficient, and satisfying for the people who use them. HCAI offers a new context for the evolution of STS principles, while STS provides a rich foundation for extending the scope and impact of HCAI.

Agile systems, today's dominant work design, can be seen as a technology-driven descendant of STS. Both emphasize teams, autonomy, and adaptability. STS is broader in scope, explicitly concerned with quality of work, democracy, and organizational values, while Agile narrows its focus to speed, customer value, and rapid adaptation. HCAI extends STS and Agile by redefining the division of labor as a collaboration between humans and machines. Automation takes on repetitive and computational tasks, enabling people to focus on judgment, creativity, and problem-solving (Schmager, Pappas, & Vassilakopoulou, 2023). This deliberate balance asks: which tasks should remain human-led, which can be machine-led, and how should the two interact? Rather than deskilling, HCAI can upskill and reskill workers, allowing them to manage more complex systems with AI as a partner. In this way,

HCAI reshapes labor, job roles, and workplace dynamics by augmenting human capabilities, enriching work with higher-order tasks, and fostering adaptive, trust-based systems that preserve human agency and meaning.

Historically, STS theory advanced a bottom-up approach to organizational design, grounded in the principle of *joint optimization* - the concurrent design of social and technical subsystems to enhance both human well-being and system performance. Over time, STS thinking has expanded from its initial focus on work-group design and production systems to encompass more complex and adaptive organizational forms, including networked structures, cross-functional teams, inter-organizational ecosystems, and customer-centered modes of value creation (Winby & Mohrman, 2018; Pasmore, Winby, Mohrman, & Vanasse, 2018; Xu & Gao, 2025). HCAI represents the next stage in this evolution, reframing technology not merely as a complementary subsystem but as an *amplifier* of human cognitive, social, and creative capacities and a *catalyst* for more adaptive, resilient, and effective organizational performance. By integrating the human-centered principles of STS with the transformative potential of AI, organizations can design systems that simultaneously advance human flourishing and strategic agility in increasingly complex environments.

Valentine et al. (2017 and 2025) introduced the concept of *flash organizations* and *flash teams* as dynamically assembled, expert teams designed to perform complex, interdependent tasks through computational coordination. Their work, developed at Stanford University, integrates socio-technical systems theory with algorithmic management to explore how human expertise and digital infrastructure can jointly enable agile, scalable collaboration. In this model, the *technical system* consists of digital platforms, workflow algorithms, and AI-based coordination tools that dynamically assemble, manage, and reconfigure expert teams based on task requirements. The *social system* comprises the human professionals—often distributed—whose skills, creativity, and collaboration drive the project outcomes. The socio-technical principle of joint optimization is embodied in flash teams by ensuring that technical automation enhances, rather than constrains, human performance—balancing efficiency, adaptability, and autonomy in real time.

3.2.6 An HCAI – Socio Technical Design Approach

The Human-Centered AI – Socio-Technical Design (HCAI–STD) is an organizational architecture that intentionally integrates people, processes, information (data), and technology to achieve optimal dynamic alignment among these elements. This alignment enables high organizational performance through the continuous evolution of business models, enhanced responsiveness to customers, and adaptive capacity to meet emerging environmental demands and opportunities. The HCAI-STD approach is presented below: *Entry, sanction, and start-up*

Sanction is obtained by reviewing the HCAI- STD approach and gaining formal approval from management in the organization. In this initial phase, education and awareness of the HCAI STD approach is provided, an organizational systems scan is conducted to determine the organizational unit of analysis under design, and stakeholder identification. Mapping accountability for decisions made by the AI system and developing guidelines for user transformation are addressed. A project structure is formed. Design goals, design criteria, and an envisioning description are created.

*Research and Analysis*

The research and analysis phase entails two analysis steps. Technical system analysis and social system analysis.

*Technical system analysis*

The technical analysis focuses on (1) processes – workflows, routines, procedures, and grouping of tasks, and (2) customer touchpoints and experience. Variances are identified in both processes, customer and ecosystem analysis. A variance indicates how much actual performance or behavior deviates from expected or standard performance metrics. Identifying variances provides insight into areas where AI-driven or human-centered AI (HCAI) solutions can enhance efficiency, quality, and user experience. Design options for both user interactions and backend processes that support AI functionality are identified. Technical analysis results in creating a prototype or pilot solution for AI and HCAI opportunities identified. This may involve developing algorithms for AI. The outcome of the technical analysis typically includes a prototype or pilot solution that embodies the AI and HCAI opportunities identified.

Technical analysis steps are outlined below:

1) Define Process
2) Map and document workflows/customer journey mapping
3) Collect KPI data / compare actual performance against expected outcomes
4) Collect data on user pain points
5) Identify variances
6) Benchmark best practices
7) Evaluate automation opportunities
8) Develop business cases for automation and HCAI -cost benefit analysis
9) Design and prototype HCAI solution
10) Implement HCAI solution / monitor performance post implementation, iterate and optimize

*Social system analysis*

The social system analysis focuses on work system unit boundaries and structures, communication and coordination processes, roles, responsibilities and skills, learning/control and adaptive system, goals, metrics, feedback, analytics, experiments, culture, and behavior. This includes an assessment of communication technologies and data requirements of work systems. Essentially, work systems are defined as key unit operations in the operating model, and the structural properties of each work system are evaluated to meet specifications. Those work system properties not meeting spec become design tasks during the HCAI- STD design phase. Social system analysis is used for testing the prototypes to ensure solutions meet HCAI requirements. Social system analysis steps include the following:

1) Conduct user research – user personas, needs assessment/pain points, task analysis
2) User centered design
3) Define requirements and constraints – functional and non-functional requirements
4) Cognitive walkthroughs - evaluate the usability of the AI interface by the user's thought process.

*Design*

Design typically occurs in a design lab environment with a specific mix of management sponsors, work systems team members, cross-functional members, supervisors, ecosystem members, customers, and internal and external functional expertise. Research and analysis data are available for design lab members to align on potential AI features and functionalities that meet user needs. Customer/user journey maps identifying touchpoints, variances, and opportunities for improvement are reviewed. Initial prototypes are created. The design lab typically entails two to three two-week sprints to reach a final design.

The design lab process includes rapid iterations on prototypes, incorporating feedback from user testing sessions to refine the design. User testing is conducted by testing prototypes with users in real or simulated environments to observe interactions and gather qualitative and quantitative feedback. A data strategy is defined that includes data collection, processing, and ethical considerations related to data usage. Typically, model development occurs by building an AI model using selected algorithms and techniques, ensuring it meets the defined requirements and is capable of addressing user needs effectively.

Various system integration and testing steps occur in the design phase**. A** plan is developed for the integration of the AI system within existing workflows or platforms, ensuring compatibility with other systems and databases. Validation and verification (unit tests, integration tests, user acceptance testing) is done to ensure the system performs as intended and meets quality standards.

The primary focus is on clarity on what tasks are handled by AI and what are managed by humans. Feedback loops are established for human oversight and AI explanations. A key design task is to ensure feedback loops are designed to detect variances in operating performance goals. Variance detection and HCAI control or elimination of variances is key to organization reconfigurability. Ethical frameworks guide the development of AI systems by emphasizing fairness, accountability, transparency, and user respect.

Once a thorough process analysis is conducted and a plan to reduce variances and automate processes is completed, the next step is to design the organization to ensure effective execution. Organization design practitioners / HR typically have separate design lab sessions that specifically involves designing the organization. The initial focus is on the organizational structure and specifically unit operations/ work system(s) that align with the redesigned process. Work system boundaries, missions, and structural properties are defined. Organizational structure is designed in ensuring network cross-functional alignment. Performance metrics are established. Roles and responsibilities are defined, skills and competencies are determined and assessed, followed by training and development. Work systems are empowered with the authority to make decisions and implement changes necessary for ongoing process optimization. Reviews occur to assess team performance, the alignment of roles with the process, and the effectiveness of the organizational structure

*Implementation*

Implementation starts with developing a comprehensive deployment strategy, including scaling, training for users, and technical support. A launch plan is created where a rollout of the AI system is in phases, allowing for gradual user adoption, and monitoring performance and user engagement. User feedback mechanisms are established to gauge user satisfaction, as well as performance analytics. It's important to establish a culture of iterative development, continuously improving the AI system based on user feedback and performance data. Adaptation/reconfigurability behaviors are established to enable adjustments to the AI model and features to adapt to changing user needs or technological advancements. Provide robust documentation for users, outlining how to interact with the AI system, its capabilities, and its limitations. Offer ongoing training sessions to educate users effectively using the AI system and understanding its implications.

### 3.2.7 HCAI-Powered Operating Model

The core organizational architecture feature of the modern organization is an HCAI-powered operating model. An HCAI-powered operating model refers to a framework in which AI technologies are integrated into an organization's processes, systems, and structures to enhance performance, drive innovation, and improve decision-making (Snow & Fjeldstad, 2014). The AI-powered operating model transforms traditional business operations by embedding intelligence in processes, enabling organizations to become more competitive in a rapidly changing landscape. Key features entail machine learning, natural language processing, computer vision, and other AI techniques into core business functions, data-driven decision making, agility and adaptability, continuous learning and improvement, and enhanced customer experience.

Modern organizations must continuously adapt and evolve alongside technology, developing both organizational and technology-based capabilities that enable ongoing improvement toward strategic objectives. The Human-Centered AI (HCAI)–powered operating model underscores the importance of the continuous alignment of people, processes, structure, information/data, and technology to support agility, efficiency, and overall organizational effectiveness.

In this model, each element must not only function effectively on its own but also complement and amplify the others, forming a cohesive, optimized operating system that drives performance and strategic outcomes. The role of organization design is to create an HCAI-powered architecture that integrates these components, optimizing their alignment to enable continuous improvement, effective customer responsiveness, and adaptive capacity in response to environmental demands and opportunities.

Central to achieving this alignment and optimization are the HCAI–Socio-Technical Design (STD) processes and design methodology**,** which provide the structured approach for configuring and integrating people, processes, information, and technology into a high-performing, adaptive operating system.

### 3.2.8 Organizational Readiness for HCAI

HCAI organizational readiness refers to the preparedness of an organization to effectively implement and use HCAI principles and practices in the development, deployment, and management of AI systems. By assessing and enhancing these areas of organizational readiness, organizations can position themselves to successfully implement HCAI practices. See Table 4 for the organization readiness assessment of HCAI.

| TABLE 4 – Organization Readiness Assessment of HCAI | | |
|---|---|---|
| # | **Readiness Area** | **Description** |
| 1. | **Long-Term Vision** | An effective strategy for HCAI should be part of a broader organizational vision that aligns technology initiatives with business goals and social responsibility. This perspective helps prioritize HCAI in the organization's strategic agenda. |
| 2. | **Policy and Governance** | Clear policies and governance structures must be in place to guide the ethical use of AI technologies. This includes addressing issues related to data privacy, security, accountability, and compliance with regulatory frameworks. |
| 3. | **Infrastructure and Resources** | Organizational readiness requires adequate technological infrastructure, including tools for user research, design, prototyping, and data management. Sufficient financial and human resources must also be allocated to support HCAI initiatives |
| 4. | **Processes and Methodologies** | The organization should have established processes and methodologies that align with HCAI principles. This may include frameworks for iterative design, user feedback mechanisms, and ethical review protocols. |
| 5. | **Stakeholder Engagement** | Readiness involves identifying and engaging various stakeholders, including users, domain experts, ethicists, and technical professionals. Their input is crucial for ensuring that HCAI systems meet diverse needs and constraints. |
| 6. | **Skill Set and Knowledge** | Employees should possess or be willing to develop the necessary skills related to HCAI, including understanding user experience, ethics in AI, design thinking, and interdisciplinary collaboration. Training and capacity-building programs may be essential for enhancing these competencies |
| 7. | **Cultural Alignment** | The organization must foster a culture that values ethical considerations, innovation, and user-centricity. Leadership support for HCAI principles and a commitment to ethical AI practices are vital for creating an environment conducive to HCAI initiatives. |
| 8. | **Commitment to Evaluation** | Organizations need a commitment to assessing the impact of HCAI systems on users and society. This involves both establishing metrics for success and integrating feedback loops to learn and adapt over time. |

## 4. HCAI-MM: Maturity Stages

The HCAI maturity model serves as a framework for organizations to evaluate their current HCAI practices and identify areas for improvement as they seek to adopt more human-centered approaches in their AI initiatives. Figure 1 provides a high-level description and objectives of the HCAI Maturity Model. By striving to progress through the stages of maturity (see Table 5), organizations can ensure that their AI systems are not only technically advanced but also aligned with human needs and ethical standards (Accenture, 2022).

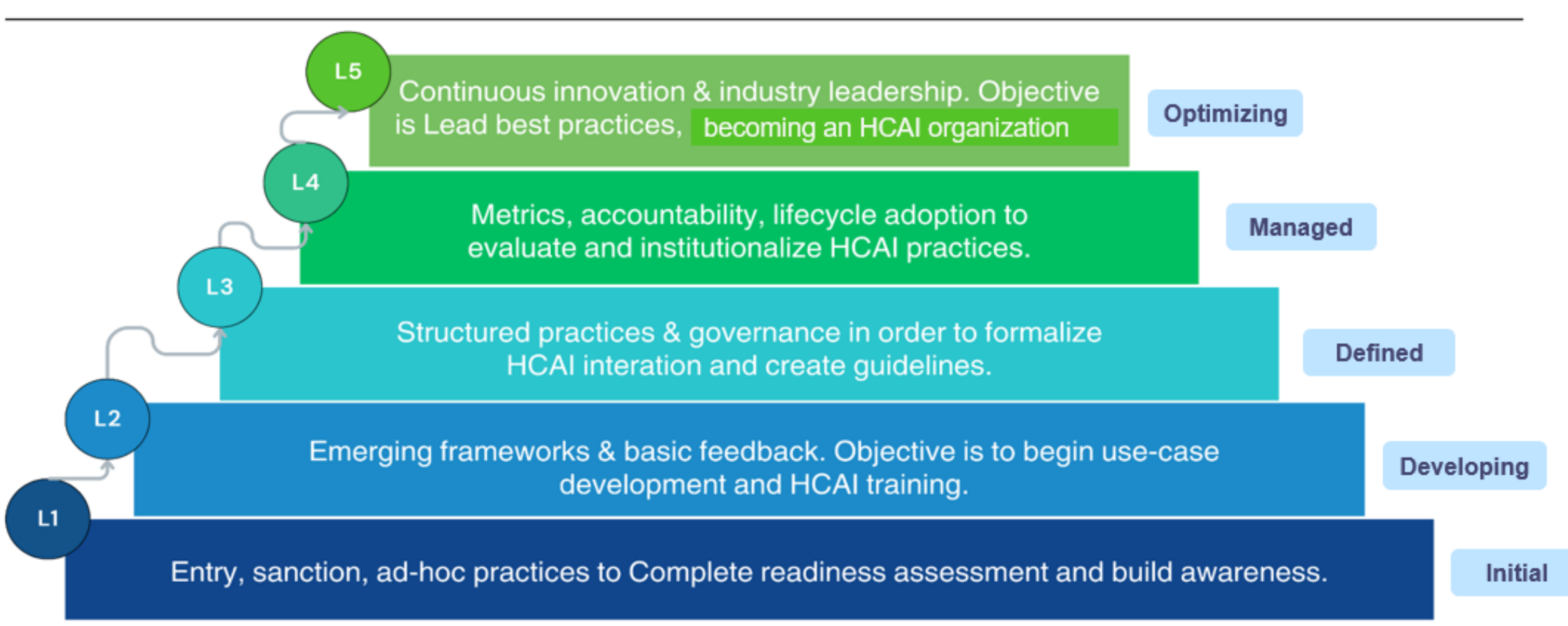

Figure 1: High-level description and objectives of HCAI Maturity Model Stages

**TABLE 5 – Stages in HCAI-MM**

| Maturity Level | Description | Characteristics | Assessment Criteria | Objectives |
|---|---|---|---|---|
| **Level 1: Initial**<br><br>Entry, sanction, and start-up design phase in progress<br><br>The organization is primarily focused on establishing a basic understanding and some ad-hoc HCAI practices | • Building sanction and resource preparation for transformation process<br>• Establishing ad-hoc practices and increase awareness of human-centered principles.<br>• Isolated efforts to address human factors in AI. | • Aligning sanction and resources for HCAI transformation<br>• AI projects initiated without user input.<br>• Limited understanding of HCAI principles. AI initiatives are generally unstructured and reactive. | • Readiness assessment in progress<br>• Entry, sanction, start-up phase in progress<br>• No formal processes for user engagement<br>• Minimal or no stakeholder feedback mechanisms<br>• Low awareness of HCAI<br>• Isolated efforts to address human factors in AI. | • Complete readiness assessment<br>• Complete entry, sanction, and startup<br>• Begin building awareness of HCAI<br>• Communicate the importance of HCAI |
| **Level 2: Developing**<br><br>Research and analysis providing awareness of opportunities and constraints<br><br>Developing an understanding and structuring HCAI practices<br><br>Begin to implement HCAI frameworks and processes | • Growing understanding of human-centered design;<br>• A structured HCAI methodology and program are in place | • User interviews or surveys conducted for select projects<br>• Basic usability testing implemented.<br>• Early collaboration between disciplines<br>• Ad-hoc training and education programs on HCAI<br>• External HCAI design standards are being adopted. | • Technical and social systems analysis is in progress.<br>• Some processes to gather user feedback<br>• Limited consideration of ethical implications<br>• Early integration of HCAI design in AI projects.<br>• Implementation of usability testing and basic feedback loops. | • Research and analysis design phase completed<br>• Foster a culture of learning around HCAI<br>• Begin documenting AI projects with a focus on understanding user needs and ethical implications.<br>• Starting the adoption of use cases. |
| **Level 3: Defined**<br>Design phase prototyping in progress.<br>Formalization of HCAI practices.<br>Actively employ and continuously refine HCAI practice | • Implementing, testing and adjusting HCAI approaches and methods<br>• Established HCAI design processes | • HCAI governance established<br>• Organizational HCAI design guidelines published<br>• Proactive HCAI-related training launched<br>• Structured deliberations in design lab<br>• Structured initiatives to incorporate user input in processes | • Structured feedback mechanisms in place<br>• Standardized processes for ethical review / governance.<br>• Formal processes for HCAI integrated into AI lifecycle<br>• Standardized practices for ensuring ethical alignment<br>• Work system team units have all structured properties in place | • Design Phase completed: prototyping, testing, and on-going HCAI iterations.<br>• Create consistent processes for involving stakeholders and users in AI system design.<br>• Develop policies and guidelines to address ethical considerations in AI. |
| **Level 4: Managed**<br>HCAI implementation stablished with widespread diffusion in organization.<br>Quantitative HCAI metrics and institutionalizing HCAI.<br>Organizations are recognized for HCAI competence, focusing on performance, accountability, and societal impact. | • Implementation with examples of HCAI strategies across the organization<br>• HCAI metrics in place. | • HCAI defined in organization strategy<br>• Continuous user involvement throughout lifecycle.<br>• Rigor HCAI governance established<br>• Data-driven decision-making based on user feedback<br>• Design lab deliberations yielding changes in work design and innovation.<br>• Comprehensive training programs on HCAI | • KPIs established for HCAI and ethical compliance.<br>• Metrics collected and acted on.<br>• Use of data-driven metrics to assess HCAI quality.<br>• Continuous monitoring of user feedback and trust in AI systems. | • Continuous HCAI implementation and iterations.<br>• Ensure all AI projects align with established HCAI policies and ethical standards.<br>• Evaluate the impact of AI systems on users and society regularly. |
| **Level 5: Optimizing**<br>Continuous improvement and innovation. Leading change/reconfigurability at all org levels through HCAI.<br>Becoming an HCAI organization.<br>Continuous performance improvement and positive experience. | • Advanced HCAI practices with a culture of continuous improvement and innovation.<br>• Evidence of continuous adaptive reconfigurability. | • HCAI is an integral part of organizational and business strategy.<br>• Active user community engaged for feedback and co-design.<br>• HCAI becomes part of the organizational culture and best practices<br>• The organization is recognized as a leader in HCAI practices within the industry. | • Regularly updated AI systems based on user insights.<br>• Formalized training programs on human-centered design.<br>• Advanced HCAI practices are embedded across the organization.<br>• HCAI becomes part of the performance dashboard (monitoring & tracking).<br>• Proactive risk management | • Identity as an HCAI organization.<br>• Proactively advocate for ethical AI development within the industry and regulatory landscape.<br>• Serve as best practice for other organizations in HCAI practice |

4.1 Key Practices across HCAI-MM 1 – 5 Levels

A number of key HCAI Practices are consistently accelerated across the five levels of the HCAI-MM, which are listed below.

- **Governance:** As organizations advance through the levels, governance structures become more formalized and integral to the AI development process, ensuring adherence to human-centered principles.
- **Stakeholder Engagement**: Progression includes increasing emphasis on engaging a diverse range of stakeholders, including end-users, ethicists, and domain experts in the AI development lifecycle. Design lab environment usage increases, providing the forum to increase social capital and cross-functional empowerment.
- **Training and Education**: At higher maturity levels, continuous training becomes a staple, with initiatives aimed at keeping the workforce updated on the latest HCAI practices and ethical considerations.

- **Impact Assessment**: Organizations should implement mechanisms for assessing the human and organizational impact of their AI systems, with an increasing focus on measurement as they mature.

## 4.2 HCAI-MM Metrics

Evaluating progress in HCAI maturity involves the use of various metrics across different maturity levels. These metrics help organizations gauge their effectiveness in implementing HCAI principles. Table 6 provides some key metrics categorized by different HCAI maturity levels, along with tools that help in measuring those metrics:

| **Table 6 – Some Key Metrics for HCAI-MM** | | | |
|---|---|---|---|
| **Maturity Level** | **Metric** | **Description** | **Tool Example** |
| **Level 1: Initial**<br>Awareness and Ad-hoc Practices. | Entry, sanction, start-up design phase. | Percentage completed entry, sanction, and start-up design phase. | Entry, sanction, and start-up methods and tranformations. |
| | Awareness and Training. | Percentage of stakeholders trained on HCAI principles and practices. | Surveys or training assessment platforms can be used to measure awareness levels among staff. |
| | Compliance with Ethical Guidelines. | Number of compliance checks against established ethical standards for AI development. | Ethical guidelines frameworks, such as those from the ACM SIGAI or the IEEE Global Initiative on Ethics of Autonomous and Intelligent Systems. |
| **Level 2: Developing** | Research and Analysis Design Phase. | Progress/percentage of completed analysis steps. | HCAI- STD Research and Analysis tools |
| | Fairness Metrics | Measures of disparity in outcomes across different demographic groups for AI systems (e.g., demographic parity, equal opportunity). | IBM's AI Fairness 360 can be used to assess fairness metrics for machine learning models. |
| | User Feedback Mechanisms. | Frequency and effectiveness of user feedback collection and analysis. | Feedback collection tools (e.g., Google Forms, SurveyMonkey) used to gather user experiences and iteratively improve AI systems based on that input. |
| **Level 3: Defined**<br>Formalization of HCAI<br>Organizations at this level begin to implement frameworks and processes to ensure HCAI principles are being adhered. | Design Phase -Prototype | Percentage of completion | HCAI-STD Design tools |
| | Model Interpretability. | The degree to which end-users understand AI system predictions (measured using user surveys). | Google's What-If Tool can provide insights into model behavior, and subsequent surveys can assess user understanding of these outputs. |
| | Usability Testing Results. | Metrics from usability tests, such as task success rate, time on task, and user satisfaction scores. | Usability testing platforms (e.g., User Testing or Optimal Workshop) can help design and evaluate tests. |
| **4 Level 4: Managed**<br>Quantitative HCAI Metrics.<br>Institutionalizing HCAI | Implementation Design Phase. | Percentage of implementation steps completion. | HCAI-STD Implementation phase tools. |
| | Ethical Compliance Audit Scores | Frequency and results of audits examining ethical compliance across AI projects. | Internal audit frameworks or external consultancy reports that evaluate adherence to ethical guidelines. |
| | Social Impact Assessment Scores. | Measures evaluating the broader social impacts of AI systems, including their influence on diversity and inclusion. | Algorithmic Impact Assessment tools, which provide structured approaches to evaluate the societal implications of AI deployment. |
| **5. Level 5: Optimizing**<br>Performance metrics in steady state.<br>Continuous Improvement and Innovation.<br>Becoming an HCAI organization. | Steady state: systems performing to requirements. | Percentage of steady state performance metrics performing to requirements. | Continuous integration /continuous deployment (CI/CD) tools that track changes and updates can provide performance metrics over time. |
| | Continuous Improvement Metrics. | Number of iterative product changes made based on user feedback and ethical evaluations. | |
| | Stakeholder Engagement | Levels of engagement with stakeholders, including users, community members, and advocacy groups. | Stakeholder engagement platforms (e.g., engagement surveys, social media analytics) to measure public sentiment and involvement. |

.

## 5. HCAI-MM: Practice

This section outlines steps for advancing the maturity model toward more sophisticated, responsible, and effective AI practices, which include tools and techniques, governance and oversight, challenges, and continuous learning and adaptation of practices.

### 5.1 Tools and Techniques

Table 7 below explores a range of tools and techniques designed to facilitate the implementation of the HCAI-MM. By leveraging quantitative and qualitative methodologies, practitioners can systematically identify gaps in their current systems and prioritize areas for improvement, which fosters a design process that aligns with HCAI principles.

| Table 7 - HCAI- MM - Tools and Techniques | | |
|---|---|---|
| **TOOL** | **PURPOSE** | **TECHNIQUES** |
| **Assessment Frameworks and Surveys** | To evaluate the current state of HCAI practices within the organization. | Custom assessment frameworks, maturity assessment surveys, and questionnaires that cover all dimensions of the HCAI-MM. These can be developed internally or through external consultation. |
| **User Research Techniques** | To gather insights about users' needs, experiences, and challenges. | **Interviews**: Conducting one-on-one discussions with users or stakeholders to understand their perspectives.<br>**Surveys/Questionnaires**: Distributing structured questions to gather quantitative data from a larger audience.<br>**Usability Testing**: Observing users as they interact with AI systems to identify pain points and areas for improvement.<br>**Focus Groups**: Facilitating discussions among a group of users to explore their experiences and expectations. |
| **Design Thinking Workshops** | To foster creativity and collaboration among interdisciplinary teams. | Facilitated workshops that encompass the five stages of design thinking (empathize, define, ideate, prototype, and test). Using tools like HCD (Human-Centered Design) methods, storyboarding, and role-playing can help generate innovative ideas and solutions |
| **Collaboration Platforms** | To enhance communication and collaboration among team members from various disciplines. | Platforms like Slack, Microsoft Teams, and Trello can be used to manage projects, share insights, and facilitate ongoing collaboration. |
| **Design Lab Deliberations** | Provides a structured process and design environment for cross-unit (function, ecosystem, customer) | Design Lab space for large group deliberation – digital/AI information processing tools and facilitation. |
| | problem solving and decision making. | |
| **Prototyping and Iteration Tools** | To visualize design concepts and test ideas quickly. | Wireframing and prototyping tools such as Figma, Sketch, Adobe XD, or Axure can help in creating low-fidelity and high-fidelity prototypes of AI interfaces and interactions. |
| **Ethical Frameworks and Guidelines** | To ensure ethical considerations are integrated into the AI development process. | Reference frameworks and guidelines like the Ethical Guidelines for Trustworthy AI (provided by the EU), organizational ethics boards, and checklists for ethical AI practices can guide the development process. |
| **Impact Assessment Metrics** | To measure the effectiveness and impact of AI systems on users and stakeholders. | Custom metrics and KPIs (Key Performance Indicators) tailored to the specific context, user satisfaction scores, accessibility audits, and performance metrics related to AI outcomes |
| **Training and Capacity-Building Programs** | To build skills and competencies related to HCAI practices within the organization. | Workshops, online courses, and seminars on relevant topics like user-centered design, ethics in AI, and collaborative methodologies. |
| **Feedback Loops and Iterative Improvement** | To create a continuous improvement cycle for AI systems. | Establishing mechanisms to collect ongoing user feedback, conducting regular retrospective meetings, and utilizing agile methodologies to incorporate changes based on user insights |
| **Stakeholder and Community Engagement Initiatives** | To ensure broader societal considerations and community needs are addressed. | Community workshops, public forums, and partnerships with local organizations or advocacy groups to co-create solutions and engage with affected populations. |

Below is a sampling of tools used by leading companies and sources employing HCAI. These tools collectively offer organizations a means to assess different dimensions of HCAI maturity.

#### 5.1.1 IBM's AI Fairness 360

- *Overview*: This is an open-source toolkit designed to help detect and mitigate bias in machine learning models. It provides a comprehensive suite of algorithms, metrics, and visualizations to assess fairness (Bellamy, Day, Hind, Hoffman, Houde, Kannan & Zhang, 2018).

- *Features*: Includes bias detection metrics (e.g., statistical parity difference, disparate impact), algorithms for mitigating bias (e.g., re-weighting), and documentation to guide users through evaluating their models.

#### 5.1.2 IBM's AI Explainability 360

- *Overview*: This toolkit provides a suite of algorithms and techniques that help users understand AI model predictions, catering to both technical and non-technical stakeholders (Google AI, 2019).
- *Features*: Offers local and global explanation methods, such as LIME (Local Interpretable Model-agnostic Explanations), SHAP (SHapley Additive exPlanations), and counterfactual explanations. It aims to foster transparency in AI applications.

#### 5.1.3 Google's What-If Tool

- *Overview:* An interactive visualization tool that enables users to analyze machine learning models without coding (Google AI.,2019). It supports the examination of model behaviors and fairness (Google AI, 2019).
- *Features*: Users can visualize model performance, analyze how changing input features affects predictions, and compare models side by side. It also facilitates the exploration of counterfactuals to understand decision-making processes.

#### 5.1.4 Microsoft Fairness Dashboard

- *Overview*: A part of Azure Machine Learning, this tool helps detect and mitigate bias in machine learning models (Mishra, Agarwal, & Mitchell, 2022).
- *Features*: Offers fairness assessments through visualizations and metrics, providing insights into potential biases affecting model predictions and helping to highlight areas for improvement.

#### 5.1.5 H2O.ai's Driverless AI

- *Overview*: An automated machine learning (AutoML) platform that claims to enhance model interpretability and accountability (H2O.ai. (n.d.).
- *Features*: Provides a set of interpretability tools and visualizations to help stakeholders understand model predictions, feature importance, and performance, facilitating a better understanding of how AI decisions are made.

#### 5.1.6 DeepAI's GPT-3 Ethics Model

- *Overview*: While not a traditional tool, it involves applying ethical considerations in the development and deployment of language models (DeepAI. (n.d.).
- *Features:* Guidelines on how to assess language model usage, focusing on ethical issues such as fairness, context-awareness, and potential impacts on marginalized groups.

#### 5.1.7 ACM SIGAI's Ethical Guidelines

- *Overview*: While not software tools, these guidelines provide principles for ethical AI development and can be used in assessment frameworks (ACM SIGAI, 2019)
- *Features*: Encourages examination of ethical considerations, stakeholder impact, and the social implications of AI systems.

#### 5.1.8 Algorithmic Impact Assessment (AIA) Tools

- *Overview*: Various AIA tools are designed to help organizations evaluate the social impact of their AI initiatives before deployment **(**Data & Society Research Institute, 2019)
- *Examples*: Tools such as the "Algorithmic Accountability Policy Toolkit" developed by the Data and Society Research Institute focus on guiding organizations through self-assessment protocols to consider ethical, legal, and policy implications.

## 5.2 HCAI Governance and Oversight

HCAI Governance and Oversight refers to the set of principles, policies, and practices designed to ensure that the development and deployment of AI systems prioritize human well-being, rights, and dignity. It encompasses a holistic approach to managing AI, focusing on transparency, accountability, fairness, and safety, with the ultimate goal of fostering trust in AI technologies among stakeholders, including users, regulators, and society at large.

### 5.2.1 Establishing ethics boards and audit mechanisms

Establishing a Human-Centered AI Ethics Board is a pivotal step for organizations seeking to ensure that their AI systems are developed and deployed in alignment with ethical standards that prioritize human well-being. This board should comprise a diverse group of stakeholders, including ethicists, technologists, social scientists, legal experts, and representatives from affected communities. By fostering a multidisciplinary approach, the ethics board can critically assess AI projects from various perspectives, addressing ethical concerns such as bias, transparency, and the potential societal impact of AI technologies. Clear terms of reference should be defined, outlining the board's responsibilities, including the review of AI initiatives before deployment, the development of ethical guidelines, and the provision of recommendations for best practices. Regular engagement with external experts and community stakeholders can enhance the board's effectiveness, ensuring that diverse voices and concerns are considered in the decision-making process.

### 5.2.2 Implementing Audit Mechanisms for Accountability

In parallel with the establishment of an ethics board, implementing robust audit mechanisms is essential for ensuring accountability in human-centered AI practices. These audit mechanisms should involve regular evaluations of AI systems to assess compliance with ethical guidelines, performance metrics, and regulatory requirements. Auditing can include both internal evaluations conducted by the organization itself and external reviews performed by independent third parties. Key components of the audit process may involve scrutinizing data sources for bias, assessing algorithmic transparency, and evaluating the impact of AI deployments on users and communities. Furthermore, audit findings should be transparently reported, fostering trust and demonstrating commitment to ethical practices throughout the AI lifecycle. Such mechanisms not only help identify areas for improvement and mitigate risks associated with AI technologies but also reinforce an organizational culture centered on responsibility, ethical considerations, and stakeholder engagement in the deployment of AI solutions.

### 5.2.3 HCAI Governance and Oversight Structures

Many companies are beginning to install HCAI Governance and Oversight structures. Below are examples of HCAI Governance and Oversight structures that companies are implementing.

**AI Ethics Boards or Committees:**

- *Structure***:** These boards are typically composed of a diverse group of experts, including ethicists, legal professionals, social scientists, data scientists, and business leaders. Some companies also involve external advisors or representatives from civil society.
- *Responsibilities***:**
    - Developing and updating AI ethics principles and guidelines.
    - Reviewing AI projects to assess potential ethical risks and impacts.
    - Providing recommendations on the design, development, and deployment of AI systems.

- Overseeing the implementation of AI ethics policies and procedures.
  - Investigating and addressing ethical complaints or incidents related to AI.
- Examples:
  - Microsoft**:** Has an AI Ethics Board that reviews and provides guidance on AI development projects.
  - Google**:** Has established an AI Ethics Board, although its composition and role have evolved over time.
  - Accenture**:** Has an AI Ethics Committee that advises on AI development and deployment.

**AI Ethics Guidelines and Policies:**

- *Content***:** Companies are developing comprehensive AI ethics guidelines that address issues such as fairness, accountability, transparency, privacy, and human oversight. These guidelines often articulate the company's values and principles related to AI.
- *Implementation***:** These guidelines are typically integrated into the AI development lifecycle, including requirements for data collection, model training, testing, and deployment.
- Examples:
  - Google's AI Principles**:** Focuses on AI's benefits, fairness, and accountability.
  - Microsoft's AI Principles**:** Encompass fairness, reliability and safety, privacy and security, inclusiveness, transparency, and accountability.
  - IBM's AI Ethics Principles**:** Emphasizes purpose, transparency, skill, and responsibility.
  - Salesforce's AI Ethics Board**:** Sets AI ethics guidelines.

**AI Risk Assessments and Audits:**

- *Process***:** Companies are implementing risk assessment frameworks to identify and evaluate the potential ethical, social, and legal risks associated with their AI systems.
- *Tools***:** These assessments often involve a combination of quantitative and qualitative methods. They might include impact assessments, bias detection tools, and fairness metrics.
- *Audits***:** Some companies are also conducting regular AI audits to assess the effectiveness of their governance structures and ensure compliance with ethical guidelines.
- *Examples***:**
  - Many companies**:** Are using internal tools and/or partnering with external auditors to assess the fairness and bias in their AI models.
  - Companies with sensitive applications**:** May use external audits to review the processes by which they developed their AI system.

**Data Governance and Privacy Frameworks:**

- *Focus***:** Ensuring the responsible collection, use, and storage of data used to train and operate AI systems.
- *Practice***:** Implementing data privacy policies, data anonymization techniques, and data access controls.
- *Example***:**
  - Companies using data for AI**:** Implementing policies to comply with regulations such as GDPR and CCPA to protect user data and privacy.

**Human Oversight and Explainability:**

- *Emphasis***:** On ensuring humans can understand how AI systems make decisions and retain control over critical decisions.

- *Practices:* Implementing methods for explainable AI (XAI), such as model interpretability tools and techniques. Establishing clear lines of accountability for AI-driven decisions.
- *Examples:*
  - Companies in sensitive industries**:** such as Healthcare or Finance, are developing methods to ensure AI decisions are transparent and are reviewed by human experts.
  - Companies providing AI systems to customers**:** are ensuring the systems are easy to understand and use.

**Training and Education:**

- *Content***:** Companies are investing in training and education programs to raise awareness of AI ethics among their employees, especially those involved in AI development and deployment.
- *Examples:*
  - Many companies**:** Are providing training on AI ethics for their employees to ensure that they understand their role in ensuring the responsible use of AI.

**Stakeholder Engagement:**

- *Activities***:** Some companies are actively engaging with stakeholders, including customers, employees, and the public, to gather feedback on their AI systems and address ethical concerns.
- *Examples***:**
  - Companies launching new AI systems**:** May invite feedback from customers and experts on the use of the systems and create a public forum to answer questions.

**Important Considerations:**

- *Context Matters***:** HCAI governance structures need to be tailored to the specific context of the company, its industry, and the types of AI systems it develops and deploys.
- *Continuous Improvement***:** AI ethics is an evolving field, and governance structures must be regularly updated to address new challenges and incorporate best practices.
- *Enforcement***:** Governance structures need to be backed by effective enforcement mechanisms to ensure compliance with ethical principles and policies.
- *Collaboration:* Collaboration between different stakeholders is essential to ensure the responsible development and deployment of AI. This includes collaboration between different departments within a company, as well as collaboration between companies, researchers, policymakers, and civil society organizations.

Many companies are actively working to establish HCAI governance and oversight structures. These structures vary in their specific details, but they generally aim to ensure that AI systems are developed and deployed in a way that aligns with human values, promotes fairness, protects privacy, and enhances human well-being.

## 5.5 HCAI-MM: Case Studies

To illustrate the effectiveness of the HCAI Maturity Model, the following two case studies from diverse industries – healthcare and technology – demonstrate how organizations have applied the HCAI-MM to achieve improvements in their use of AI.

### Case Study 1: Human-Centered AI in Healthcare Operations – Mayo Clinic

At Level 2 of HCAI maturity, the Mayo Clinic implemented a human-centered AI system to optimize clinical workflow scheduling and reduce clinician burnout (Verghese, Shah, & Harrington, 2023). The AI tool used natural language processing to analyze patient visit data and predict appointment durations. Rather than automating scheduling entirely, the system provided adaptive recommendations that clinical teams could review and modify. The human-centered design process included iterative feedback from physicians, nurses, and administrative staff, ensuring transparency and trust in AI outputs. This collaborative approach improved scheduling accuracy by 18% and reduced administrative burden, while preserving clinician decision authority—hallmarks of Level 2 maturity where humans remain central in AI-informed processes.

**Case Study 2: Human-Centered AI in Talent Management – IBM's AI-Augmented HR System**

IBM's human resources organization introduced a human-centered AI platform to assist managers with talent retention and internal mobility decisions (Boudreau, & Jesuthasan, 2022). At Level 2 of the HCAI maturity model, the system provided predictive insights on employee attrition risk, complemented by explainable dashboards that allowed HR partners to question or adjust AI-driven suggestions. The organization established ethical review checkpoints and employee feedback loops to ensure fairness and interpretability. Rather than replacing human judgment, the AI augmented decision-making and fostered data-informed conversations between managers and employees. This integration represented an intermediate level of maturity—structured, explainable, and participatory—enabling IBM to align AI with human values and organizational trust.

In essence, these case studies demonstrate how different organizations are using the Human-Centered AI Maturity Model to improve their use of AI and achieve their goals.

## 6. Challenges and future work

### 6.1 Challenges of HCAI practice within organizations

Organizations may encounter the following challenges as they seek to navigate the complexities of integrating HCAI into their AI development and deployment processes.

#### 6.1.2 Cultural Resistance

One of the primary challenges organizations face is cultural resistance to change. Many organizations have established practices and mindsets around AI development that prioritize efficiency and technological advancement over human-centered considerations. Shifting this culture involves overcoming ingrained habits, biases, and a lack of awareness regarding the importance of ethical considerations in AI. Engaging employees, fostering buy-in from leadership, and promoting awareness of HCAI principles are essential but may require significant time and effort.

#### 6.1.3 Resource Allocation

Implementing the HCAI-MM necessitates the allocation of resources, including time, personnel, and financial investment, which can be a barrier for many organizations. Establishing ethics boards, conducting audits, and ensuring compliance with human-centered guidelines require dedicated teams with diverse skill sets, including ethics, social science, and technology. Additionally, organizations may need to invest in training and education to enhance their capabilities in HCAI principles. This resource allocation may detract from other pressing business priorities, leading to conflicts regarding budgeting and strategic focus.

#### 6.1.4 Measurement and Evaluation Difficulties

Another significant challenge lies in developing effective metrics and evaluation frameworks to assess progress within the HCAI-MM. Unlike traditional performance metrics that often focus on efficiency and profit, human-centered outcomes are inherently more subjective and complex to measure. Organizations may struggle to define clear indicators of success in HCAI practices, such as user satisfaction, fairness, or societal impact. The absence of standardized measurement tools can hinder the ability to assess maturity levels effectively and to provide actionable insights for improvement.

#### 6.1.5 Regulatory and Compliance Challenges

As organizations adopt the HCAI-MM, they must navigate a complex landscape of existing and emerging regulations related to AI ethics, data privacy, and accountability. Understanding and complying with diverse regulatory requirements across jurisdictions can be daunting, and organizations may face challenges in aligning their practices with varying legal standards. Furthermore, as technology evolves, regulators are continuously working to develop new frameworks, creating a dynamic environment where compliance requires ongoing adaptation and vigilance.

#### 6.1.6 Integration with Existing Systems and Processes

Successfully integrating the HCAI-MM into existing systems and processes can be challenging. Organizations often have legacy systems and established workflows that may not easily accommodate the necessary changes to support HCAI principles. This integration may require significant modifications to development pipelines, decision-making frameworks, and organizational structures. Ensuring that HCAI considerations are woven into the fabric of AI development may necessitate iterative changes, which can be time-consuming and complex.

### 6.2 Continuous learning and adaptation of practices

Continuous learning and adaptation of HCAI practices refer to the ongoing process of improving AI systems and their related practices based on feedback, new insights, evolving user needs, and changes in social, ethical, and technological landscapes. By emphasizing continuous learning and adaptation, organizations involved in AI development can better align their practices with the HCAI principles and overcome the challenges in HCAI practices. This approach emphasizes the importance of flexibility, responsiveness, and iterative development in creating AI solutions. Below are some key components and aspects of continuous learning and adaptation in HCAI.

1) **Feedback Loops**: Establish mechanisms for capturing and analyzing feedback from users and stakeholders throughout the AI lifecycle. This can include user satisfaction surveys, focus groups, and community engagement initiatives. Regularly soliciting input helps organizations understand how AI systems impact users and can highlight areas in need of improvement.
2) **Iterative Design and Development**: Embrace an iterative approach to product development that allows for rapid prototyping, testing, and refining of AI systems. This iterative cycle fosters experimentation and agility, enabling teams to adapt quickly to new information and user feedback. Techniques such as Agile methodologies or Design Thinking can be particularly useful in facilitating this process.
3) **Learning from Data**: Leverage data analytics and machine learning to continuously evaluate the performance of AI systems. Analyze usage patterns, outcomes, and user interactions to identify trends and root causes of issues. This data-driven approach allows organizations to make informed decisions about necessary adjustments and improvements.
4) **Ethical Reflection**: Engage in regular ethical reflection and review of AI systems. Assess the ethical implications of deploying AI in real-world contexts, considering potential unintended consequences, bias, and discrimination. Continuous ethical evaluation ensures that the systems remain aligned with human values and societal norms.
5) **Professional Development and Training**: Invest in ongoing training and development for team members involved in AI design and implementation. This can include workshops, seminars, and access to resources focused on HCAI principles, ethics, and technical skills. Continuous upskilling enables teams to stay informed about best practices and emerging issues in the field.
6) **Benchmarking and Best Practices**: Identify and learn from industry benchmarks, case studies, and best practices related to HCAI. Organizations can adapt successful strategies from peers and thought leaders to enhance their own practices. This comparative analysis can provide insights into effective methods for user engagement and ethical AI practices.

7) **Adaptation to Regulatory Changes**: Monitor and adapt to new regulations, standards, and best practices related to AI and ethics. As laws evolve, organizations must be proactive in modifying their policies and practices to ensure compliance and uphold ethical standards.

8) **Interdisciplinary Collaboration**: Foster collaboration among diverse disciplines, including technology, psychology, sociology, and ethics. Engaging experts from various fields can provide holistic perspectives on how AI systems should be designed, deployed, and evaluated. Collaborative efforts can also help identify and mitigate biases and assumptions inherent in AI development.

9) **Community Involvement**: Actively involve users and community members in the ongoing evaluation of AI systems. Engaging with the broader community not only enhances transparency but can also provide crucial insights into how AI applications affect different groups of people. This involvement can take several forms, including user advisory panels or community forums.

## 7. Conclusion

The evolution of AI within organizations requires a structured framework to align AI technologies with human values and organizational design. Without such a framework, organizations risk ethical dilemmas, operational misalignments, and discord among people, processes, and technology. The Human-Centered AI Maturity Model (HCAI-MM) addresses this need by providing a framework for organizations to evaluate and enhance their HCAI practices. It outlines five stages of maturity, guiding organizations in assessing current practices, identifying gaps, and improving capabilities. With a focus on HCAI guiding principles, socio-technical methodologies, and metrics for assessing maturity, the HCAI-MM promotes responsible AI development, alignment among stakeholders, and integration of technology with human and organizational needs. By offering a structured methodology for continuous learning and adaptation, the HCAI-MM model empowers organizations to mitigate ethical risks, optimize performance, and cultivate resilient ecosystems where technology, people, and processes operate in harmony—ultimately fostering impactful and human-centered AI solutions.

**Acknowledgements:**

In creating this paper, we have drawn on the experience and thinking of colleagues to whom we are indebted, especially Terry Carroll, Peter Gaarn, Mark Govers, Paul Hower, Dean Myers, Bill Passmore, Bert Painter, and Chris Worley.